\documentclass{appolb}
\usepackage{epsfig}

\begin{document}
\pagestyle{plain}
\title{Non-congruent phase transitions in cosmic matter and in the laboratory}

\author{Igor Iosilevskiy
\address{Joint Institute for High Temperature of RAS,
125412 Moscow, Russia}
\address{Moscow Institute of Physics \& Technology 
, 141700 Moscow, Russia}
} \maketitle

\begin{abstract}
Non-congruence appears to be the most general form of phase transition
in cosmic matter and in the laboratory. In terrestrial applications
non-congruence means coexistence of phases with different chemical
composition in systems consisting of two (or more) chemical elements. 
It is just the case for all phase transitions in
high-temperature chemically reactive mixtures, which are typical for
uranium-bearing compounds in many nuclear energy devices, both
contemporary and perspective. 
As for cosmic matter, most of real and hypothetical phase transitions 
without nuclear reactions, i.e., those in the interiors of giant planets 
(solar and extrasolar), those in brown dwarfs and other sub-stellar objects, 
as well as in the outer crust of compact stars, are very plausible candidates 
for such type of phase transformations. 
Two exotic phase transitions, the gas-liquid phase transition in dense nuclear 
matter and the quark-hadron transition occuring in the interior of compact 
stars as well as in high-energy heavy-ion collisions are under
discussion as the most extreme example of hypothetical non-congruence
for phase transformations in High Energy Density Matter.
\end{abstract}

\PACS{PACS numbers:  64.60.Bd; 95.30.Qd; 97.10.-q; 25.75.Nq}


\section{Introduction}

The term \emph{non-congruent phase transition} (NCPT) denotes the situation of 
phase coexistence of two (or more) phases with different chemical compositions.
This is a rather evident definition for the case of phase transitions (PT)
in most of terrestrial applications (see below) and in
astrophysical applications, where nuclear transformations, including
$\beta$-decay, are negligible: PTs in planetary science and outer
crust of compact stars etc. 
The nuclear composition in such situations is conserved and there is no 
problem with the selection of systems, which fulfill the condition of a NCPT. 
The situation is more complicated under extreme conditions like in the 
interiors of compact stars and in remnants of supernova explosions, where 
nuclear transformations are close to equilibrium. 
The problem of the NCPT relevance is even more complicated in exotic 
situations with equilibrium hadron decay and quark deconfinement in interiors 
of strange (hybrid) stars and in the hypothetical quark-hadron (QH) phase 
transition in ultrarelativistic heavy-ion collisions at RHIC, LHC, FAIR
and NICA. 
Hence the study of non-congruent phase transitions in typical terrestrial 
applications could be a useful base for understanding the relevance for such 
type of phase transitions in exotic situations like interiors of compact stars,
supernova explosions, and in the hydrodynamic expansion of a fireball formed in
heavy-ion collisions.

\section{
General features of non-congruent phase transitions in chemically
reactive plasmas}

Phase equilibrium in chemically reactive non-ideal plasmas of two or
more chemical elements differs fundamentally from the case of
ordinary phase equilibrium like, for example, the Van der Waals PT in
substances with fixed chemical compositions (stoichiometry). 
Phase transitions in chemically reactive mixtures, including those in
high-temperature uranium-bearing compounds, are typical for many
nuclear energy devices both contemporary \cite{UO2-book} and
perspective \cite{Ievlev,MONO-80}. 
The basic feature of such two-phase systems is their non-congruency, i.e. 
their ability to vary stoichiometries of coexisting phases without violating 
the stoichiometry for the whole two-phase mixture. 
Non-congruency changes significantly the properties of all phase transitions 
in such systems, namely: 

(\textbf{A}) - The significant impact of the \emph{phase
transformation dynamics}, i.e. of the strong dependence of the phase
transition parameters on the rapidity of the transition. 
This dependence is of primary importance in experiments with fast surface
evaporation of condensed samples under the powerful laser heating or
electron-beam energy deposition. 
The strong competition between diffusion and thermal conductivity processes 
determines the parameters of such non-congruent evaporation; 

(\textbf{B}) - The \emph{phase transition thermodynamics} becomes more 
complicated. 
The essential changes include the scale of two-phase boundaries in
extensive thermodynamic variables (say $P$-$\rho$ etc) and even in
topology of all two-phase boundaries in the space of intensive
thermodynamic variables, as well as properties and even nature of
the singular points (critical point first of all) and appearance of
additional end-points in NCPT. One of the most remarkable
consequences of the non-congruency is the change of the general form of the 
two-phase boundary in the pressure-temperature plane (see Fig.~\ref{fig-1}
below). 
A two-dimensional "banana-like" region appears in the NCPT instead
of the well-known one-dimensional $P-T$ saturation curve for ordinary
(congruent) PTs. 
A next remarkable property for a NCPT is that isothermal and isobaric 
crossovers of the two-phase region are no longer coincide. 
The isothermal NCPT starts and finishes at different pressures, while the 
isobaric NCPT starts and finishes at different temperatures \cite{IosUO2}. 
This property is crucial for the interpretation of the NCPT relevance in 
the physics of compact stars and high-energy heavy-ion collisions.

\section{Conditions of joint phase, chemical and ionization equilibrium}

\subsection{Equilibrium between macroscopic phases with neutral species}

\subsubsection{Maxwell conditions}

Phase equilibrium conditions for two macroscopic phases are well
known for the case when coexisting phases consist of arbitrary
mixtures of neutral species with equilibrium chemical reactions. In
accordance with chemical thermodynamics laws these conditions
include conditions of equilibrium heat and impulse exchange
(equality for pressures and temperatures: $P'= P''$, $T'= T''$) and
conditions of equilibrium matter exchange. The latter conditions
have two variants for systems consisting of two or more chemical
elements. The first one corresponds to \emph{partial} equilibrium
for exchange of matter with \emph{fixed chemical composition}. This
condition is equivalent to the well-known Maxwell "equal squares"
construction for pressure-volume dependence in the case when both
coexisting phases can be described by unique thermal equation of
state (EOS) $P(V,T)$. For example, it is so for Van der Waals
(gas-liquid) phase transition.

More general is the well-known "double tangent" construction for two
free energies, $F'(V,T,x)$ and $F''(V,T,x)$ when both coexisting
phases are described by different EOS-s. For example, it is so for
crystal-fluid phase transition. In both the variants the final
equilibrium condition corresponds to equality of Gibbs free energies
of coexisting phases with fixed chemical composition:

\begin{equation}
T' = T''~,~~~ P' = P''~,~~~ G'(T, \rho', x)=G''(T, \rho'', x)
\end{equation}

This form of phase equilibrium condition is often noted as "Maxwell
condition" in astrophysical literature (for example
\cite{NS_Ioffe}).

\subsubsection{Gibbs conditions}

The second variant corresponds to the \emph{total} equilibrium in
mean-phase matter exchange, i.e. equilibrium for exchange by each
species with varying chemical composition of coexisting phases $(x'
\neq x'')$, but without violation of total chemical composition of
whole two-phase system. This variant leads to not one, but several
separate equalities for partial quantities - chemical potentials
$\mu_i$ ($i$=1,2,. . . k - all species) at $T'=T''$ and $P'=P''$

\begin{equation}
\mu_i'(T, \rho', x') = \mu_i''(T, \rho'', x'')~,~~~ \alpha x'+ (1
-\alpha)x''= x
\end{equation}

In terrestrial applications this form of phase equilibrium
conditions corresponds exactly to the definition of non-congruent
phase transition. For the case of equilibrium chemical reactions in
each phase total number of equalities for chemical potentials is
decreased to reduced number of equalities for chemical potentials of
basic (independent) species. For example, it is two basic units,
oxygen and uranium chemical potentials $\mu_\texttt{O}$ and
$\mu_\texttt{U}$, in the case of equilibrium uranium-oxygen mixture
(see below). In astrophysical application the form (2) is well known
under the name "Gibbs conditions". The problem is that this form is
applied there to charged species, but not only to neutral ones (see
below).

\subsection{
Phase equilibrium of macroscopic phases in presence of charged
species (Gibbs - Guggenheim conditions)}

Phase equilibrium conditions for macroscopic phases with charged
species are more complicated. There are two basic points. The first
one is that electroneutrality conditions are added for both phases
in (1, 2). Maxwell conditions (1) are still valid for Gibbs free
energies, $G'$ and $G''$, of \emph{electroneutral} phases with
chemical and ionization equilibrium inside. As for the Gibbs
conditions (2), the point is that besides electroneutrality
restrictions two additional quantities appear in description of
coexisting phases and, correspondingly, in equilibrium conditions as
additional independent variables. It is average electrostatic
potentials, $\varphi'(r)$ and $\varphi''(r)$ \cite{Guggen} (see for
example \cite{IoChi00}). As a result, a remarkable feature of any
Coulomb system is the existence of two versions of chemical
potential,  $\mu_i$ and  $\tilde{\mu}_i$. The (ordinary) chemical
potential, $\mu_i({n_k},T)$, is presumed to be a local parameter
depending on local density, temperature and composition. The new
(generalized) \emph{electro-chemical potential}  $\tilde{\mu}_i$ is
not local parameter. It strongly depends on non-local sources of
influence, such as total charge disbalance including surface dipole,
other surface properties etc. In \emph{uniform} Coulomb system
$\tilde{\mu}_i$ is equal to the sum of $\mu_i({n_k},T)$ and average
(bulk) electrostatic potential, $\varphi$, which is \emph{presumed}
to be \emph{uniform} also.

\begin{equation}
\tilde{\mu}_i =\mu_i(\{n(\textbf{r})\},T(\textbf{r})) + Z_i e
\varphi(\textbf{r})
\end{equation}

For each charged specie in Coulomb system the values of its chemical
potentials in coexisting phases, $\mu_i'$ and  $\mu_i''$, must not
be equal under conditions of phase equilibrium. It is namely the
electro-chemical potential, to have the same values in coexisting
phases at phase equilibrium:

\begin{equation}
 \mu_i'(\{n'\},T) \neq \mu_i''(\{n''\},T)~,~~~
 \tilde{\mu}_i'(\{n'\},T) = \tilde{\mu}_i''(\{n''\},T)
\end{equation}

This form of phase equilibrium conditions (3, 4) we will refer below
as Gibbs-Guggenheim conditions. Equalities (3, 4) being combined
with the electroneutrality conditions leads to remarkable feature of
any equilibrium Coulomb system, namely: every phase boundary in such
system is accompanied, as a rule, by a finite gap in the average
electrostatic potential through the phase interface
\cite{IoChi92,IoChi00}.

\begin{equation}
 \bigtriangleup\varphi \equiv \varphi''(r \rightarrow +\infty)
 - \varphi'(r \rightarrow - \infty) = (\mu_e'' - \mu_e')(e)^{-1}
 = (\mu_i' -\mu_i'')(Ze)^{-1}
\end{equation}

In contrast to the work function this inter-phase (Galvani)
potential drop $\bigtriangleup\varphi$  represents a thermodynamic
quantity, which does depend on temperature and chemical composition
only and does not depend on surface properties. This gap tends to
zero at the critical point of gas-liquid phase transition. The
zero-temperature limit of this drop (along the coexistence curve)
can be considered as an individual thermo-electrophysical
coefficient of any material. The value of discussed potential drop
could be directly calculated by numerical modeling of phase
transitions in the Coulomb system when both the coexisting phases
being explicitly simulated \cite{IoChi00}.

It should be stressed that any phase transition in plasmas of one
chemical element, for example evaporation in metals, must be
forced-congruent in spite the fact that one (or both) coexisting
phases being composed of two basic units: ions and electrons (all
other species being their equilibrium bound complexes). It is
electroneutrality conditions in both (macroscopic) phases that make
this coexistence thermodynamically one-dimensional. On the contrary,
this system became two-dimensional (and all phase transitions became
non-congruent) just at the moment when we relax electroneutrality
conditions in both phases and allow equilibrium mean-phase exchange
by charged species also. This is just the case in so-called "mixed
phase" scenarios (see below).

\subsection{Mesoscopic scenarios for phase equilibrium
("mixed phase" concept)}

There exists very popular and widely accepted scenario for phase
transition, which differs essentially from the both described above
ones. The basic idea, which was claimed in \cite{Mix-83} and
developed in \cite{Glend-92} and many other papers (for example
\cite{Glen_Bla}), is that in many astrophysical situations a highly
dispersive, uniform and heterogeneous mixture of charged
micro-fragments of one phase in oppositely charged "see" of another
one (charged emulsion) may be more thermodynamically favorable (i.e.
not metastable, as it is in most of terrestrial applications (mist,
foam etc.) but stable) than standard (Maxwell) form of
forced-congruent coexistence of two electroneutral macroscopic
phases.

The simplest form of mixed-phase equilibrium conditions is
equivalent exactly to equations (2) as if all charged species are
equilibrated in mean-phase exchange as well as neutrals species. In
this simplest approximation all thermodynamic loss in such charged
emulsion due to Coulomb energy of charge separation and positive
contributions of surface tension are neglected.

More sophisticated form of discussed mesoscopic scenario for phase
coexistence ("structured mixed phase", see for example
\cite{Jap-Voskr}) takes into account mentioned above thermodynamic
loss due to surface tension and charge separation. It leads to
existence of optimal size, form and charge for micro-fragments of
both mixed phases in mentioned above charged emulsion ("pasta
plasma"). The question of degree of equivalence for 'structured
mixed phase' and non-congruent PT is open. See discussion below.

\section{Non-congruent evaporation in the uranium-oxygen system}

\begin{figure}[thb]
\includegraphics[width=0.8\textwidth]{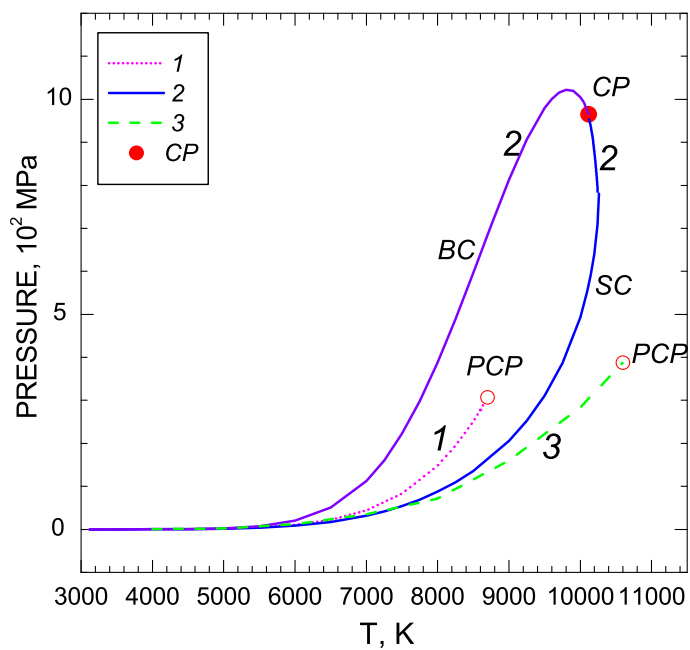}
\caption{
Pressure-temperature diagram for non-congruent evaporation
in chemically reacting U-O plasma (O/U = 2.0) \cite{IosUO2}.
\emph{1,2}-- calculations via EOS \cite{IosUO2}: \emph{1}--
forced-congruent coexistence (Maxwell conditions (1)), \emph{PCP} --
pseudo-critical point.
\emph{2-2}-- boundaries of the two-phase region via total
(non-congruent) equilibrium Eqs. (2)-(5): $BC$-- boiling curve (bubble
point), $SC$-- saturation curve (dew point), $CP$-- true critical
point; \emph{3}-- gas-liquid coexistence curve calculated via
previous EOS of UO$_2$ \cite{Fischer}.
\label{fig-1}}
\end{figure}

Development of wide-range equation of state(EOS) for uranium and
uranium-bearing compounds with taking into account all phase
transformations in such systems, was the subject of multi-annual
theoretical study \cite{MONO-80,UO2-book}. The physics of phase
transitions in uranium dioxide (UO$_{2\pm x}$) is of primary
importance for prediction of behavior of nuclear reactors during
hypothetical severe accidents \cite{UO2-book}. In set of the works
\cite{INTAS, IosUO2,IoUO2-03} an adequate theoretical model of
non-congruent evaporation in U-O system was developed. The basic
point of the model is the description of both coexisting fluid
phases (liquid and vapor) in a uniform manner, as equilibrium
multi-component strongly interacting (non-ideal) mixtures of atoms,
molecules, molecular and atomic ions, and electrons as well
("chemical picture", see e.g. \cite{MONO-80}). Chemical reactions
and ionization as well as the parameters of phase equilibrium have
been calculated self-consistently by taking into account all
significant non-ideality corrections (strong Coulomb interaction,
intensive short-range repulsion and attraction) within modified
version of thermodynamic theory. Details of the adopted
approximations are described elsewhere \cite{IosUO2,UO2-book}. The
fluid model (common for liquid and vapor phases) has been applied
for self-consistent calculations of non-congruent phase coexistence
within the wide range of temperature and pressure ($T\leq$ 20 kK,
$P\leq$ 2 GPa) including the vicinity of the true critical point of
non-congruent PT. The basic point of these calculations is that the
Gibbs conditions (2) have been used for all neutral species in both
phases, while the Gibbs - Guggenheim conditions (3,4) have been used
for all charged species, the conditions, which have been actually
violated in all previous studies of evaporation in U-O system (for
example \cite{Fischer}). The pressure-temperature phase diagram for
non-congruent evaporation is shown in Fig. \ref{fig-1} as the most important
result for present discussion.

\section{
General nature of non-congruent phase coexistence in compounds and
chemical mixtures}

Mentioned above long-time study of non-congruent phase equilibrium
in U-O system  \cite{IosUO2,IoUO2-03} indicates that this
type of phase transformation is not as infrequent at high
temperatures as it was seen before. The main conclusion drawn from
above results could be formulated in following statement:

{\emph Any phase transition in equilibrium system containing two
or more chemical elements must be non-congruent in general.
Congruent phase transitions in such systems arise as exception only.}

This statement seems to be in evident contradiction with our
everyday experience because one knows very many examples of PT-s in
compounds of two (or more) chemical elements, for example, in simple
water and other substances (H$_2$O, CO$_2$, NH$_3$ etc), where
parameters of PT are studied exhaustively and nobody ever heard
about non-congruence and banana-like $P$-$T$ diagrams. Nevertheless,
there is no contradiction. Gas-liquid PT in all these compounds are
exceptions. All these PT-s are indeed congruent in \emph{room
conditions} because all of them conserve mono-molecular composition
through the evaporation (H$_2$O $\rightleftharpoons$ H$_2$O) and
there is no any degree of freedom for two-phase system to change
stoichiometry in liquid and/or vapor phases. But situation is
absolutely different for PT-s in these compounds in planetary
conditions ($T\sim$ 10-20 kK, $P\sim$ 1-10 Mbars). Expected
nomenclature of PT-s in such conditions is very abundant (see for
example \cite{Matts}) while all discussed compounds are no more
mono-molecular. Our present knowledge of parameters for these PT-s
is very poor \cite{H2O}. But qualitatively the main statement of
present work is that any phase transition in these compounds in
planetary conditions must be non-congruent, i.e. all $P$-$T$ (or
$\mu$-$T$) boundaries for phase transitions must be two-dimensional
regions instead of ordinary one-dimensional curves \cite{IoUO2-03}.

Generally, the expected examples of non-congruent phase transitions
in terrestrial applications are inter alia:
\begin{itemize}
\item Uranium- and plutonium-bearing compounds: (PuO$_{2\pm x}$, UC, UN
etc),
\item Evaporation in other oxides (for example, in SiO$_2$)
\item Evaporation in hydrides of metals (for example, in LiH)
\item Evaporation in ionic liquids and molten salts: (for example, in
NaCl)
\item Evaporation in metallic alloys
\item Phase transitions in "dusty" and colloid plasmas: (Coulomb system
of macro- and micro-ions with charge $q_M = + Z, q_m = \pm 1$)
\end{itemize}

\section{Non-Congruence in cosmic matter}

\subsection{Ordinary situations}

There exist many candidates for such type of phase transitions in
cosmic matter without nuclear transformations:
\begin{itemize}
\item Hypothetical plasma- and dissociation-driven phase transitions in
mixture H$_2$ + He (+ H$_2$O + NH$_3$ + CH$_4$) in interiors of
giant planets (Jupiter, Saturn, Neptune etc), in brown dwarfs and in
extra-solar planets \cite{IoUO2-03},
\item Phase transitions in isentropically released products of strong
shock compression of lunar ground (SiO$_2$ + FeO + Al$_2$O$_3$ + CaO
+ ...) under huge impact after natural (meteorite) or artificial
(LCROSS mission) bombarding
\item Crystallization and ionic demixing in interiors of white dwarfs,
\item Crystallization and ionic demixing in outer envelopes of compact
stars (for example \cite{Horowitz}).
\end{itemize}

\subsection{Non-congruence in exotic situations}

Relevance of non-congruent scenario for phase transition in exotic
situations is not transparent. There exist many phase transitions,
which could be considered as candidates for such transformations.
Two groups of them will be commented here as the first ones:

(I) Gas-liquid (Van der Waals-like) phase transition in dense
nuclear matter of equilibrium mixture of $p, n, e$ and nuclei
$\{N(A, Z)\}$. Here $\{N(A, Z)\}$ is equilibrium ensemble of all
possible bound complexes from $Z$ protons and $(A - Z)$ neutrons
(see \cite{Typel} and reference therein). Several variants may be
considered: with and without electrons, electroneutrality and
Coulomb interaction, and with and without $\beta$-equilibrium).

(II) Hypothetical phase transition(s) in the vicinity of quark
deconfinement boundary at high temperature and with very complicated
nomenclature of hypothetical phase transformation at relatively low
temperature (for example \cite{Glen_Bla}).

\subsubsection{Gas-liquid phase transition}

(I.a) Gas-liquid (GL) phase transition (PT) in equilibrium mixture
$\{p, n, N(A, Z)\}$ with no electrons, no electroneutrality, no
Coulomb repulsion (for example \cite{Typel,GL-FIAS}). The system is
equivalent to chemically reacting mixture of two chemical elements.
The symmetry parameter $Y$ - is independent variable. It is
equivalent to stoichiometry (chemical composition) in ordinary
chemical mixtures. Hence, this GLPT is non-congruent in
non-symmetric case ($Y \neq$ 0.5) and congruent (i.e. aseotropic) in
symmetric case ($Y$ = 0.5).

(I.b) Ordinary GLPT (with macroscopic coexisting phases) in the same
mixture $\{p, n, e, N(A, Z)\}$ with electrons, Coulomb interaction
and electroneutrality and with $\beta$-equilibrium. The system is
equivalent to \emph{one-dimensional} (thermodynamically) system like
partially ionized and dissociated hydrogen $\{p, e, H, H_2, H^-,
H_2^ + \dots {\rm etc.}\}$ (Gibbs-Guggenheim conditions (3-5)).
Hence, this GLPT is forced-congruent (VdW-like).

(I.c) GLPT in the same mixture as in (I.b) $\{p, n, e, N(A, Z)\}$ in
frames of simplest mesoscopic scenarios (simple "mixed phase"). No
local electroneutrality, only global one. (Gibbs conditions (2)).
This system is again equivalent to two-component (two-element)
chemically reacting terrestrial mixture. Hence, this GLPT is
non-congruent in general. $P-T$ and $\mu-T$ phase boundary must be
two-dimensional banana-like region instead of ordinary (VdW-like)
saturation curve.

(I.d) GLPT in the same mixture as in (I.c) in frames of advanced
mesoscopic scenarios ("structured mixed phase" - "pasta plasmas").
The most complicated situation. This system is not equivalent to any
terrestrial analog. Problem of congruence for such GLPT should be
analyzed separately.

\subsubsection{Quark-hadron phase transition}

(II.a) Quark-hadron (QH) phase equilibrium (PT) between macroscopic
quark-gluon and hadron phases is one-dimensional (thermodynamically)
system. Phase transitions must obey to the Gibbs-Guggenheim
conditions (3-5). Hence this variant of QHPT is equivalent to
congruent PT, i.e. $P$-$T$ and $\mu$-$T$ phase boundaries must be
one-dimensional curves rather than two-dimensional stripes. It
should be stressed that this variant of QHPT is not equivalent to
VdW-like PT (like case I.b) by two reasons. First, this variant of
QHPT is much closer to entropic type of PT (i.e. decreasing $P$-$T$
coexisting curve, small density gap etc.) than to enthalpic one like
VdW-PT (i.e. increasing $P$-$T$ coexisting curve, large density gap
etc.) \cite{Io_Enc-00} \cite{Io_Enc-04}. Second, presently
considering versions of QHPT are described by separate analytic
EOS-s for quark and hadron phases. Hence, there is no reason to
expect appearance of critical point in such descriptions like it is
in the case of crystal-fluid phase transition in terrestrial physics
\cite{LaLi}.

(II.b) QHPT in the same combination as in (II.a) in frames of
simplest \emph{mesoscopic} scenarios (simple "mixed phase").  No
local electroneutrality, only global one. Gibbs conditions (2) are
valid for all species, charged and neutral. Quark-hadron phase
transition via "mixed-phase" scenario has the main features of
non-congruent phase transitions: Isothermal transitions through the
two-phase region starts and finishes at \emph{different pressures}
(and at \emph{different partial chemical potentials}). This system
is equivalent to two-dimensional (thermodynamically) system. Hence,
this version of QHPT is non-congruent in general. $P$-$T$ and
$\mu$-$T$ phase boundaries are two-dimensional stripes rather than
one-dimensional curves.

(II.c) QHPT in the same combination as in (II.a) in frames of
advanced mesoscopic scenario: "structured mixed phase" - "pasta
plasma" (for example \cite{Jap-Voskr}). The most complicated
situation. This system is not equivalent to any terrestrial analog.
Problem of congruence or non-congruence for such QHPT should be
analyzed separately.

\section{Acknowledgements}

The work was supported by the grants: INTAS-93-66, CRDF MO-0110, ISTC
3755 and by the RAS Scientific Program "Physics of matter under extreme
conditions" and by the MIPT Education Center "Physics of high energy
density matter". We acknowledge especially David Blaschke for great
support and very useful discussions.



\begin{thebibliography}{99}

\bibitem{Fischer}
E.A. Fischer, \emph{Journ. of Nucl. Sci. and Eng.} \textbf{101},
97 (1989);
G. Gigli,  M. Guido, G. De Maria, \emph{Journ. of Nucl. Mater.}
\textbf{98}, 35 (1981).

\bibitem{GL-FIAS}
L. Satarov, M. Dmitriev, I. Mishustin, \emph{Physics of Atomic Nuclei},
\textbf{72}, 1390 (2009).

\bibitem{Glend-92}
N.K. Glendenning, \emph{Phys. Rev. D}, \textbf{46}, 1274 (1992);
\emph{Compact stars}, Springer, Berlin (1997).

\bibitem{Glen_Bla}
D. Blaschke,  N.K. Glendenning,  A. Sedrakian,
\emph{Physics of Neutron Star Interiors}, Lecture Notes in Physics,
Vol. \textbf{578}, Springer, Berlin (2001).

\bibitem{Guggen}
E. Guggenheim, \emph{J. Phys. Chem.} \textbf{33}, 842 (1929);
\emph{Modern Thermodynamics by the Methods of Willard Gibbs},
Methuen, London (1933).

\bibitem{H2O}
 M. French, T. Mattsson, N. Nettelmann, R. Redmer, \emph{Phys. Rev.
 B}, \textbf{79}, 054107 (2009).

\bibitem{Horowitz}
 C.J. Horowitz, D.K. Berry, E.F. Brown, \emph{Phys. Rev. E}, \textbf{75},
066101 (2007).

\bibitem{Ievlev}
 V. Ievlev, \emph{Bulletin of Russ. Acad. Sci.}, \textbf{6}, 24 (1977);
 \emph{Rocket engines and energy converters based on gas-core nuclear
 reactor}, Ed. A. Koroteev, Machinery Publishing, Moscow (2002)
 (in Russian).

\bibitem{INTAS}
INTAS-93-66 Final Reports, European Commission, JRC, Institute for
Transuranium Elements (Karlsruhe) 1997, 1999.

\bibitem{IoChi92}
I. Iosilevski,  A. Chigvintsev, in \emph{Physics of Non-ideal
Plasmas}, Eds. W. Ebeling, A. F\"orster, R. Radtke, Teubner,
Stuttgart-Leipzig (1992), p.87; arXiv:physics/0612113.

\bibitem{IoChi00}
I. Iosilevski and  A. Chigvintsev, \emph{Journal de Physique}
\textbf{IV 10}, Pr5, 451 (2000).


\bibitem{Io_Enc-00}
 I. Iosilevskiy and A. Starostin, in \emph{Encyclopedia of Low-Temperature
Plasma Physics}, Ed. V. Fortov, Nauka, Moscow (2000), p. 327-339 (in
Russian)

\bibitem{Io_Enc-04}
 I. Iosilevskiy, in \emph{Encyclopedia of Low-Temp. Plasma Physics},
Suppl. III-1, Eds. A. Starostin and I. Iosilevskiy, Fizmatlit,
Moscow (2004), p. 349-428 (in Russian).




\bibitem{IosUO2}
I. Iosilevski,  G. Hyland,  E. Yakub,  C. Ronchi, \emph{Transactions
of the Am. Nucl. Society}, \textbf{81}, 122 (1999); \emph{Int.
Journ. Thermophys.} \textbf{22}, 1253 (2001).

\bibitem{IoUO2-03}
I. Iosilevskiy,  V. Gryaznov,  E. Yakub,  C. Ronchi,  V. Fortov,
\emph{Contrib. Plasma Phys}. \textbf{43}, 316 (2003).

\bibitem{Jap-Voskr}
 T. Maruyama, T. Tatsumi, D.N. Voskresensky, T. Tanigawa, S. Chiba,
\emph{Phys. Rev. C} \textbf{72}, 015802 (2005).

\bibitem{LaLi}
 L.D. Landau, E.M. Lifshitz, \emph{Statistical Physics},
 Fizmatlit, Moscow (1995).

\bibitem{Matts}
 T. Mattsson, M. Desjarlais, \emph{Phys. Rev. Lett.} \textbf{97}, 017801
(2006).

\bibitem{MONO-80}
V. Gryaznov,  I. Iosilevskiy,  V. Fortov et al.
\emph{Thermophysics of gas-core nuclear reactor}, Ed. V. Ievlev,
Atomizdat, Moscow (1980) (in Russian).

\bibitem{Mix-83}
D.G. Ravenhall,  C.J. Pethick and J.R. Wilson, \emph{Phys. Rev. Lett.}
\textbf{50}, 2066 (1983).

\bibitem{NS_Ioffe}
P. Haensel,  A.Y. Potekhin,  D.G. Yakovlev,
\emph{Neutron Stars: Equation of State and Structure},
Springer, New York (2007).

\bibitem{Typel}
S. Typel,  G. R\"opke, T. Kl\"ahn,  D. Blaschke, H.H. Wolter,
\emph{Phys. Rev. C} \textbf{81}, 015803 (2010); arXiv:0908.2344.

\bibitem{UO2-book}
 C. Ronchi, I. Iosilevskiy, E. Yakub,
\emph{Equation of State of Uranium Dioxide}, Springer, Berlin (2004).

\end{thebibliography}
\end{document}